\documentclass{ws-procs975x65}

\begin{document}

\title{HAWKING RADIATION FROM BLACK HOLES OF CONSTANT NEGATIVE CURVATURE VIA GRAVITATIONAL ANOMALIES}

\author{PETROS SKAMAGOULIS}

\address{Department of Physics, National Technical University of Athens,\\
 Zografou Campus GR 157 73, Athens, Greece\\
E-mail: pskam@central.ntua.gr}

\begin{abstract}
I derive the Hawking flux from black holes of constant negative
curvature and from a black hole of constant negative curvature
conformally coupled to a scalar field, using the covariant
gravitational anomalies method.
\end{abstract}


\bodymatter


\section{Introduction}

A few years ago Wilczek and collaborators proposed a new method to
derive the Hawking flux from spherical black
holes\cite{Robinson:2005pd,Iso:2006wa,Iso:2006ut}. They have proved,
using a dimensional reduction procedure, that near the event horizon
effective field theories become two-dimensional and exhibit gauge
and gravitational anomalies. They showed, in the Unruh vacuum, that
the Hawking flux emerges in order to cancel these anomalies. Recent
applications and a modification of this method can be found in
Refs.~\citen{apps:2009,Morita:2009mt}. A more technically simplified
version of the method for asymptotically flat spacetimes, using only
the covariant form of the anomalies, was suggested in
Ref.~\citen{Banerjee:2008az}. The purpose of this brief article,
which is based in Ref.~\citen{Papantonopoulos:2008wp}, is to apply
this method of covariant anomalies in asymptotically non-flat
spacetimes of non-spherical topology.


\section{Hawking Radiation from Topological Black Holes}\label{TBHsec}

Consider the four-dimensional black hole solutions of Einstein's
equations with negative cosmological constant $\Lambda=-3l^{-2}$.
These asymptotically locally Anti-de Sitter (AdS) black holes are
known as topological black holes (TBHs)\cite{Vanzo:1997gw} and their
metric is
\begin{equation}
ds^{2}=-f(r)dt^{2}+\frac{1}{f(r)}dr^{2}+r^{2}\left(d\theta^{2}+\sinh^{2}\theta
d\varphi{^2}\right)\,,\quad
f(r)=-1+\frac{r^{2}}{l^2}-\frac{2\mu}{r}\,,
\label{TBH}
\end{equation}
where $l$ is the AdS radius and $\mu>-l/3\sqrt{3}$ is a constant
proportional to the black hole mass. Their topology is
$\mathbb{R}^{2}\times\Sigma$, where $\Sigma$ is a two-dimensional
manifold of constant negative curvature. The angular part of
\eref{TBH}, with $0\le\varphi<2\pi$ and $\theta\ge0$, is the line
element of the manifold $\Sigma$, which is locally isomorphic to the
hyperbolic manifold $H^{2}$ and of the form $\Sigma=H^{2}/\Gamma$,
where $\Gamma\subset O(2,1)$ is a discrete subgroup of isometries of
$H^2$. The manifold $\Sigma$ is a compact two-dimensional surface of
constant negative curvature and of genus
$\bar{\mathrm{g}}\geq2$\cite{Papantonopoulos:2008wp}. Below, I will
concentrate only on the genus $\bar{\mathrm{g}}=2$ case and deal
with the event horizon $r_{H}$, which corresponds to the outer
horizon for $-l/3\sqrt{3}<\mu<0$ and to the unique horizon for
$\mu\geq0$.

Now consider a complex scalar field $\phi(x)$ in the background of
the TBH of genus two. Its action is
\begin{equation}
S=-\frac{1}{2}\int
d^{4}x\sqrt{-g}\phi^{*}\nabla^{2}\phi+S_{\mathrm{int}}\,,
\label{actiontop}
\end{equation}
where the first term is the free part of the action and
$S_{\mathrm{int}}$ is the part which includes a mass term, potential
terms and self-interaction terms. I perform a partial wave
decomposition of $\phi$ in terms of the functions
$\mathcal{Y}_{\xi}^{m}$, which are defined as
\begin{equation}
\mathcal{Y}_{\xi}^{m}(\theta,\varphi)=\left(\frac{2\pi}{\xi\tanh(\pi\xi)}\right)^{\frac{1}{2}}
\frac{\Gamma(i\xi+\frac{1}{2})}{\Gamma(i\xi+m+\frac{1}{2})}
P_{-\frac{1}{2}+i\xi}^{m}(\cosh\theta)e^{im\varphi}\,,
\label{Yfunct}
\end{equation}
where $P_{-(1/2)+i\xi}^{m}$ are the associated Legendre functions,
$m=0, \pm1, \pm2, \pm3, \ldots$ and $\xi$ takes discrete real values
$\xi\geq0$. These functions form a complete orthonormalized set of
functions on the manifold $\Sigma$ of genus
two\cite{Papantonopoulos:2008wp,Argyres:1988ga}. Upon transforming
to the ``tortoise'' coordinate $r_{*}$ defined by $ dr/dr_{*}=f(r)$
and taking the near-horizon limit $r_{*}\rightarrow-\infty$, one
finds that the effective radial potentials for partial wave modes of
the field, the mass term and the interaction terms contain the
suppression factor $f(r(r_{*}))$ and vanish exponentially
fast\cite{Papantonopoulos:2008wp}. Keeping only dominant terms, I
find that physics near the event horizon can be effectively
described by an infinite collection of two-dimensional free massless
scalar fields, each propagating in a two-dimensional spacetime,
which is given by the $(t,r)$ part of the TBH metric of
\eref{TBH}\cite{Papantonopoulos:2008wp}.

In this reduced two-dimensional spacetime, following the ideas of
Wilczek et al.\cite{Iso:2006wa,Iso:2006ut}, I neglect the ingoing
modes in the region near the event horizon, which behave as
left-moving modes, since they can not classically affect physics
outside the horizon. Then, the effective two-dimensional theory
exhibits a gravitational anomaly in this region, in the form of the
non-conservation of the energy-momentum
tensor\cite{Iso:2006wa,Iso:2006ut}. The time component of the
covariant gravitational anomaly
is\cite{Banerjee:2008az,Papantonopoulos:2008wp}
\begin{equation}
\partial_{r}\tilde{T}^{r}_{t}=\frac{1}{96\pi}f\partial_{r}f''=\partial_{r}\tilde{N}^{r}_{t}\,,
\quad\text{where}\quad\tilde{N}^{r}_{t}=\frac{1}{96\pi}\left(
ff''-\frac{f'^{2}}{2}\right)\,,
\label{Ward}
\end{equation}
$\tilde{T}^{\mu}_{\nu}$ is the covariant energy-momentum tensor and
the prime denotes differentiation with respect to $r$. Solving this
equation with the proposed in Refs.~\citen{Iso:2006wa,Iso:2006ut}
boundary condition $\tilde{T}^{r}_{t}(r_{H})=0$, I find the
non-conserved energy-momentum tensor
\begin{equation}
\tilde{T}^{r}_{t}(r)=\tilde{N}^{r}_{t}(r)-\tilde{N}^{r}_{t}(r_{H})\,.
\label{Trt}
\end{equation}
The Hawking flux is measured at infinity, where there is no
gravitational anomaly and in
Refs.~\citen{Robinson:2005pd,Iso:2006wa,Iso:2006ut} it was given by
the conserved energy-momentum tensor. One sees that the
gravitational anomaly $\partial_{r}\tilde{N}^{r}_{t}$ vanishes at
asymptotic infinity. However, $\tilde{N}^{r}_{t}$ is non-zero there,
due to the fact that the spacetime asymptotically is AdS. These
remarks and \eref{Ward}, which can be written as
$\partial_{r}\left(\tilde{T}^{r}_{t}-\tilde{N}^{r}_{t}\right)=0$,
imply that the conserved energy-momentum tensor, or the energy flux
$F$ measured at infinity, is
\begin{equation}
F=\tilde{T}^{r}_{t}(r\rightarrow\infty)-\tilde{N}^{r}_{t}(r\rightarrow\infty)=-\tilde{N}^{r}_{t}(r_{H})
=\frac{\pi}{12}\left(\frac{f'(r_{H})}{4\pi}\right)^{2}\,.
\label{Trtcons}
\end{equation}
The energy flux $F$ has a form equivalent to blackbody radiation of
temperature $T=f'(r_{H})/4\pi$\cite{Iso:2006wa,Iso:2006ut}, which is
exactly the Hawking temperature of the TBH of
\eref{TBH}\cite{Vanzo:1997gw}. Hence, $F$ can be identified with the
Hawking flux from this TBH. However, I should note that the actual
Hawking radiation observed at infinity is calculated through the
grey-body factors, due to the curvature of the spacetime away from
the horizon\cite{Robinson:2005pd}.


\section{Hawking Radiation from a ``Hairy'' Topological Black Hole}\label{MTZsec}

Another interesting non-spherical black hole solution is a TBH
conformally coupled to a scalar field $\Psi$. This black hole
solution is known as the MTZ black hole\cite{Martinez:2004nb}
\begin{eqnarray}
ds^{2}&=&-f(r)dt^{2}+\frac{1}{f(r)}dr^{2}+r^{2}\left(d\theta^{2}
+\sinh^{2}\theta d\varphi{^2}\right)\,,\\
f(r)&=&\frac{r^{2}}{l^{2}}-\left(1+\frac{G\mu}{r}\right)^{2}\,,\qquad\Psi(r)=\sqrt{\frac{3}{4\pi
G}}\frac{G\mu}{r+G\mu}\,,
\label{MTZconf}
\end{eqnarray}
where $G$ is Newton's constant and $\mu>-l/4G$ is a constant
proportional to the black hole mass. The MTZ black hole has the same
topology as the TBHs previously analyzed. Therefore, I will
concentrate again only on the case of $\Sigma$ being a genus two
compact surface of constant negative curvature and deal with the
event horizon $r_{+}$, which is the unique horizon for $\mu\geq{0}$
and the outer horizon for $-l/4<G\mu<0$.

Considering a complex scalar field $\phi$ in the background of the
MTZ black hole of genus two, which does not interact with the scalar
field $\Psi$, the dimensional reduction procedure proceeds as in
\sref{TBHsec}. All the arguments of \sref{TBHsec} hold also in this
case\cite{Papantonopoulos:2008wp} and the conserved energy-momentum
tensor is
\begin{equation}
F=\tilde{T}^{r}_{t}(r\rightarrow\infty)-\tilde{N}^{r}_{t}(r\rightarrow\infty)=-\tilde{N}^{r}_{t}(r_{+})
=\frac{\pi}{12}\left(\frac{f'(r_{+})}{4\pi}\right)^{2}\,.
\label{TrtconsMTZ}
\end{equation}
This energy flux  has a form equivalent to blackbody radiation of
temperature $T=f'(r_{+})/4\pi$\cite{Iso:2006wa,Iso:2006ut}, which is
the Hawking temperature of the MTZ black hole. The conformally
coupled scalar field $\Psi$, which is time-independent, does not
contribute to the Hawking flux. This was proved explicitly in
Ref.~\citen{Papantonopoulos:2008wp} by showing that its action near
the horizon vanishes, due to the presence of the suppression factor
$f(r(r_{*}))$.

Conclusively, I derived the Hawking temperature and flux, without
backscattering effects, from non-spherical black holes in
asymptotically non-flat spacetimes, via the method of covariant
anomalies, and showed that an infinite set of two-dimensional scalar
fields near the horizon acts as the thermal source of this flux.



\begin{thebibliography}{99}

\bibitem{Robinson:2005pd}
  S.~P.~Robinson and F.~Wilczek,
  {\em Phys. Rev. Lett.} {\bf 95}, 011303 (2005).

\bibitem{Iso:2006wa}
  S.~Iso, H.~Umetsu and F.~Wilczek,
  {\em Phys. Rev. Lett.} {\bf 96}, 151302 (2006).

\bibitem{Iso:2006ut}
  S.~Iso, H.~Umetsu and F.~Wilczek,
  {\em Phys. Rev. D} {\bf 74}, 044017 (2006).

\bibitem{apps:2009}
  L.~Bonora, M.~Cvitan, S.~Pallua and I.~Smolic,
  {\em Phys. Rev. D} {\bf 80}, 084034 (2009);
  J.~J.~Peng and S.~Q.~Wu,
  arXiv:0906.5121 [hep-th];
  S.~W.~Wei, R.~Li, Y.~X.~Liu and J.~R.~Ren,
  {\em Eur. Phys. J. C} {\bf 65}, 281 (2010);
  R.~Becar, P.~Gonzalez, G.~Pulgar and J.~Saavedra,
  arXiv:0808.1735 [gr-qc].

\bibitem{Morita:2009mt}
  T.~Morita,
  {\em Phys. Lett. B} {\bf 677}, 88 (2009).

\bibitem{Banerjee:2008az}
  R.~Banerjee,
  {\em Int. J. Mod. Phys. D} {\bf 17}, 2539 (2009).

\bibitem{Papantonopoulos:2008wp}
  E.~Papantonopoulos and P.~Skamagoulis,
  {\em Phys. Rev. D} {\bf 79}, 084022 (2009).

\bibitem{Vanzo:1997gw}
  L.~Vanzo,
  {\em Phys. Rev. D} {\bf 56}, 6475 (1997).

\bibitem{Argyres:1988ga}
  E.~N.~Argyres, C.~G.~Papadopoulos, E.~Papantonopoulos and K.~Tamvakis,
  {\em J. Phys. A} {\bf 22}, 3577 (1989).

\bibitem{Martinez:2004nb}
  C.~Martinez, R.~Troncoso and J.~Zanelli,
  {\em Phys. Rev. D} {\bf 70}, 084035 (2004).

\end{thebibliography}
\end{document}